# Design and modeling of an ultra-compact 2x2 nanomechanical plasmonic switch

Vladimir A. Aksyuk[1,*]

[1]Center for Nanoscale Science and Technology, National Institute of Standards and Technology, 100 Bureau Dr., Gaithersburg, MD 20899, USA
[*]vladimir.aksyuk@nist.gov

**Abstract:** A 2x2 Mach-Zehnder optical switch design with a footprint of 0.5 μm x 2.5 μm using nanomechanical gap plasmon phase modulators [1] is presented. The extremely small footprint and modest optical loss are enabled by the strong phase modulation of gap plasmons in a mechanically actuated 17 nm air gap. Frequency-domain finite-element modeling at 780 nm shows that the insertion loss is ≤ 8.5 dB, the extinction ratio is > 25 dB, and crosstalk for all ports is > 24 dB. A design optimization approach and its dependence on geometrical parameters are discussed.

## Introduction

A nanomechanical gap plasmon (GP) phase modulation principle has been recently proposed and experimentally verified [1]. Such modulators were predicted to scale down to below a 1 μm² footprint without loss of performance, and applications in miniaturized optical switching were proposed. In order to realize such switches, other optical components are necessary. A 2x2 Mach Zehnder switch (schematically shown in Figure 1c) is a prototypical example, where two 3 dB couplers and input and output ports are implemented and incorporated with two arms, at least one of which is phase modulated. One possibility is to couple the plasmonic modulators to a more commonly used dielectric waveguide technology, such as silicon on insulator, where such components are routinely implemented. Low loss coupling techniques between dielectric waveguides and GPs, as well as low loss coupling to very narrow gaps have been reported [2, 3]. However, such conventional dielectric waveguides and couplers cannot be further downscaled, hindering realization of the full potential for miniaturization inherent in the plasmonic approach.

It is therefore desirable to realize such miniaturized components directly for GPs and tightly integrate them into fully plasmonic switches. Proposed in this work and numerically validated is a design for such an integrated 2x2 switch. First, two-dimensional (2D) finite-element modeling (FEM) is used for modal analysis of propagating, confined GP modes. The results are then used to derive geometrical parameters for the switch. Finally, using a full-vector three-dimensional (3D) frequency-domain FEM, the switch 3dB coupler length is further optimized and its expected performance is predicted. With full-3D verification the GP phase modulator function is found to be in agreement with [1]. Note that the π phase modulators in this work have a smaller footprint than [1], 0.5 µm² (0.25 µm x 2 µm), including the lateral air gaps on both sides. All FEM is implemented and carried out via a commercial finite-element solver. Remarkably, modest loss and high extinction ratio can be expected from such small switches.

**Switch design**

The switch model consists of two stacked, initially identical, gold parts with an air gap in between (Figure 1). Such a device is feasible to fabricate, for example, by first depositing gold, thin silicon dioxide or other dielectric sacrificial layer and then another layer of gold. The device shape can be formed by electron beam or modern high resolution stepper lithography (100 nm lines and spaces are required) with subsequent ion beam milling. The sacrificial layer, removed by an isotropic etch, forms an air gap. The device can be actuated, for example electrostatically [1, 5]. An additional structural layer (not shown) is envisioned on top of the device, as is common in multilayer surface micromachining [5]. This layer is structured in a way that only the movable modulator arm is allowed to move, while the coupler and reference arms are held rigidly in place. Following [1] we use a parabolic approximation for the deformation shape of the movable arm (vertical deformation is shown via color scale in Figure 1). While a complete switch design would necessarily include both electrostatic and mechanical actuation analysis in full detail, as well as a suitable fabrication sequence, the goal of this work is to study and verify the expected optical performance of such devices, for which a parabolic approximation of deformation is adequate.

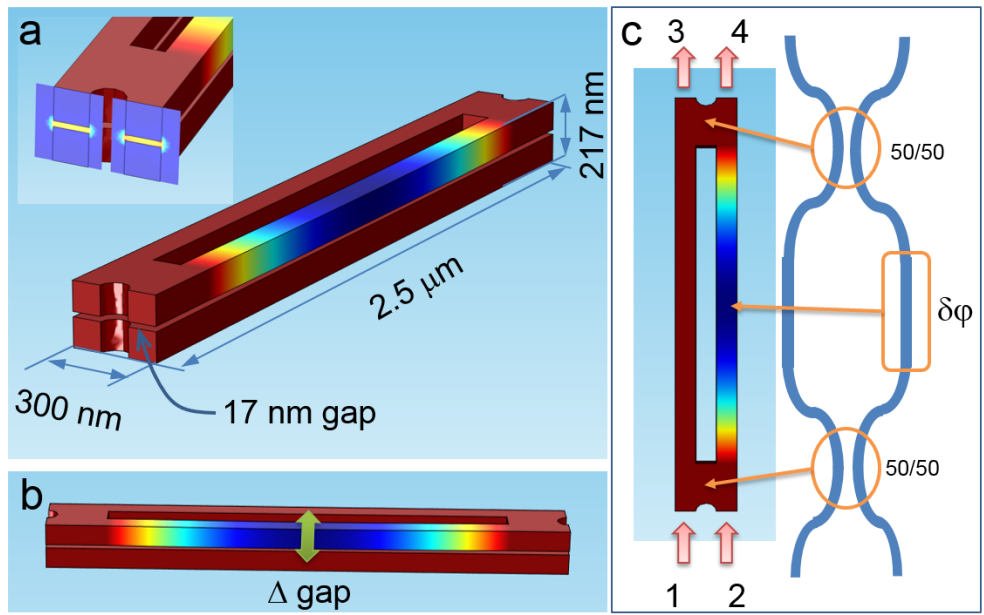

Fig. 1. Nanomechanical gap plasmon 2x2 optical switch. (a) Perspective views and dimensions. Color scale – vertical displacement of the actuated modulator arm. Inset – input/output port GP modes calculated by FEM: electric field norm. (b) Side view with the vertically actuated modulator arm (c) top view and schematic showing the input and output ports, 3 dB couplers, the modulator arm and the reference arm.

Figures 1a,b show perspective and side views of the switch with only one arm actuated down. The confined GP modes are seen in the Fig. 1a inset. As seen in Figure 1c, bottom to top, the switch consists of two input GP ports, a 3 dB (50/50) coupler region, two separate GP waveguide arms, one of which is actuated to shift the GP phase by δϕ, another 3 dB coupler and two output ports. The calculated modes for each of the two single-mode GP input ports are graphically shown in the inset to Figure 1a. The optical fields are well-confined in the gap and the effective index is approximately 2 for a 17 nm gap with 100 nm wide ports. The ports are separated by a 100 nm lateral air gap, as are the two arms of the switch.

In this work, the optical wavelength in vacuum is $\lambda_0$ = 780 nm and gold is the plasmonic metal (dielectric constant = -22.4476 + 1.36505$i$ at this frequency), so that the modulator design from [1] can be directly followed. However, due to the broadband nature of GPs, with an appropriate choice of sizes, the same type of GP switch can be made for any wavelength which is sufficiently below the surface plasmon cutoff wavelength and where no strong absorption otherwise exists in the metal of choice.

*2.1 Initial geometrical parameter estimates*

The initial gap and length for the phase modulator and reference arms are $g_0$ = 17 nm and $L_{arm}$ = 2 µm to provide π phase modulation for an approximately $\Delta gap$ = 0.28$g_0$ = 4.76 nm gap change at midpoint, with 1/e (4.3 dB) insertion loss [1]. The width of the individual arms is chosen to be much wider than $g_0$, so that the effective index is close to that of an infinitely wide GP and the phase modulation theory in [1] remain quantitatively valid. On the other hand, they should be narrow enough for the GP to remain single mode. An arm width of 100 nm satisfies these requirements. A narrower width may be possible, but may require longer arms and larger separation and result in a higher optical loss.

Arms of width 100 nm ensure that the GP is well confined and that the two arms can be reasonably close together without excessive mixing of their modes. To properly choose the inter-arm distance, the optical modes are modeled for the two arms together for several inter-arm separations. The individual GP modes form symmetric and antisymmetric combinations (Figure 2a), having slightly different effective indexes ($n_1$ and $n_2$). The index difference becomes larger with smaller separations and stronger mixing. If light is initially injected into only one of the arms, the energy becomes equally distributed between both arms when a π/2 phase difference is accumulated between the modes, requiring the travel length of $L_{3dB}^{arm}$ = $(\lambda_0/4) \cdot (n_1^{arm} - n_2^{arm})^{-1}$. This distance should be larger than the arm length. For an inter-arm separation of 100 nm, the calculated indexes are $n_1^{arm}$ = 2.066, $n_2^{arm}$ = 2.000, and $L_{3dB}^{arm}$ = 3 µm. The 100 nm width and 100 nm separation are chosen for the arms as well as for the input and output ports.

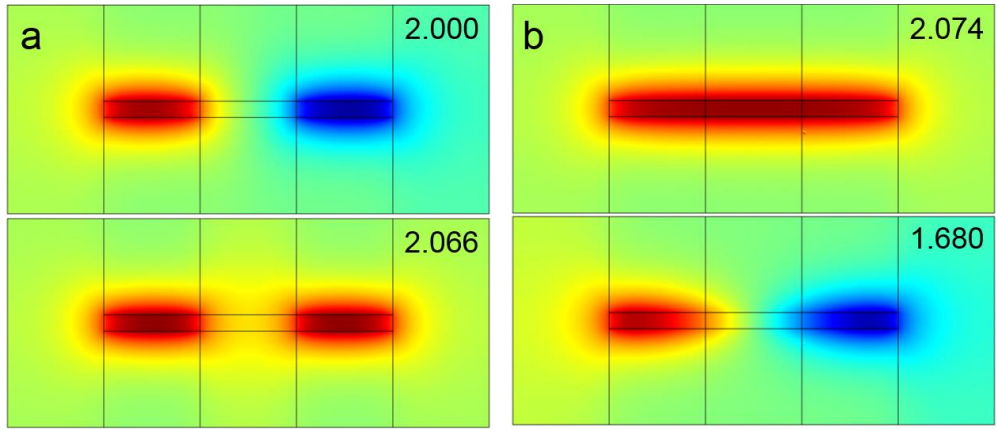

Fig. 2. Gap plasmonic mode FEM calculations. (a) Symmetric and antisymmetric combinations of single arm plasmonic modes with their effective indexes (b) Coupled first and second modes and their effective indexes. Only the real parts of the complex refractive indexes are quoted.

A similar calculation is used to design a 3 dB coupler. The modes and effective indexes are calculated for a 300 nm wide GP waveguide, that only has two modes. Visual comparison of the mode shapes in Figure 2b to Figure 2a, and taking into account their indexes, leads to the conclusion that a reasonably good mode overlap and therefore acceptable scattering and reflection losses can be expected, even for an abrupt transition between the two separate arms and the 300 nm slab. The 300 nm slab is sufficiently narrow, so that there is a sizable difference between the refractive indexes of its two modes. Given that the calculated $n_1^{slab} = 2.074$ and $n_2^{slab} = 1.680$, the slab length should be about $L_{3dB}^{slab} = (\lambda_0/4) \cdot (n_1^{slab} - n_2^{slab})^{-1} = 495$ nm to work well as a 3dB coupler.

However, given the $L_{3dB}^{arm} = 3$ μm calculated above, some coupling is expected to continue in the arms. For an unactuated device, the aim is to couple all the power from port 1 to port 4 and from port 2 to port 3. Therefore we need the total phase shift between the symmetric and antisymmetric modes accumulated by traversing the entire switch to be $\pi$ which is seen below:

$$2\pi \cdot (n_1^{slab} - n_2^{slab}) \cdot 2L^{slab}/\lambda_0 + 2\pi \cdot (n_1^{arm} - n_2^{arm}) \cdot L^{arm}/\lambda_0 = \pi$$

This results in an initial approximation for the slab length of $L^{slab} = 327$ nm.

An abrupt transition between the 3 dB coupler region and the switch arms is left in place, while a more gradual transition is made between the ports and the couplers using a 100 nm diameter semi-circular cut. These transition shapes can be further improved in the future to minimize the switch insertion loss.

*2.2 Coupler length optimization*

What follows is a full 3D finite element model implementation of the switch. First used are symmetric arms (no mechanical deformation). Calculation of the S parameters of the switch uses numerical solutions for the 2D port modes on each of the four port faces (Figure 1a inset) as boundary conditions for the 3D problem, implementing the ports numerically. The port boundaries use mode overlap integrals to model the incoming optical mode energy (e.g. on either port 1 or port 2), while absorbing the outgoing energy in the particular modes. In this implementation, the energy not matching the mode was reflected, which is reasonable for single-mode ports. The GP can then be injected through either port 1 or port 2 to calculate the transmission to port 3 (S31, S32), and port 4 (S41, S42), as well as reflection (S11, S22) and cross-coupling S12. While full complex S-parameters are calculated (including phase), only transmitted and reflected power is reported, in units of dB.

S31 and S41 are calculated as a function of the coupler length (the same for both couplers). The coupler length is defined as the distance along the length of the switch (direction normal to the port faces) from the end of the semi-circular gap between the ports (50 nm from port faces) to the start of the arms. Figure 3 shows that the highest port 4 power rejection (S41 minimum) occurs for a coupler length slightly more than 200 nm. This is somewhat shorter than the initial guess of 327 nm. Furthermore, by observing the time-harmonic oscillating optical field in the switch, it is evident that some amount of reflection occurs from the transition regions, and therefore the fields in the switch are somewhat more complicated than the simple analysis based on unidirectional travel would suggest. Only a weak reflection is observed (a small apparent oscillation of intensity of traveling waves through the switch), indicating no strong resonances and therefore broadband operation is expected, to be confirmed by a future study.

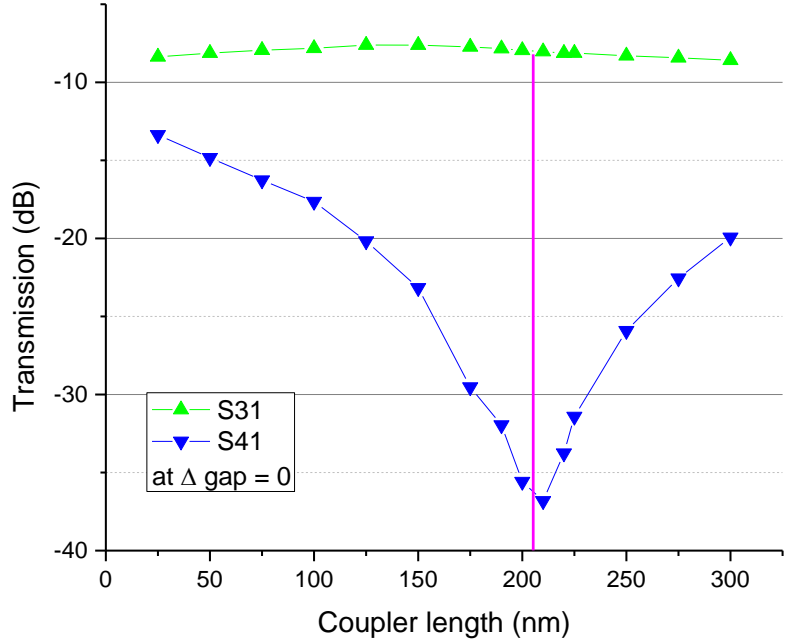

Fig. 3. Coupler length optimization in the OFF state. Optimal coupler length is near the high extinction point

Using an optimal coupler length of 200 nm, the total length of the switch is 2500 nm, including 50 nm for ports on each end. The physical width of the switch is 300 nm, however a more adequate number for the footprint is 500 nm, which is the same as the width of the model. This includes 100 nm air margins on each side to ensure that the fields have space to decay appropriately (the estimated longest 1/e power decay rate is $L_{dec}^{-1} \approx 2\,(n_{eff} - 1)^{1/2}\,k_0 = (62\text{ nm})^{-1}$). This results in a switch footprint area of 0.5 μm x 2.5 μm = 1.25 μm².

**Switch performance**

Switch functionality is modeled by deforming the FEM mesh of the modulator arm and the surrounding air to approximate actuation. The mesh elements of the arm are moved down by an explicitly prescribed displacement. The mesh in the air outside the arm is moved down by the same distance, while the deformation in the gap is linearly interpolated, such that it is zero at the gap's lower surface, so that the mesh remains continuous and is not inverted by deformation. The air mesh deformation between the arms is also linearly interpolated such that it is zero at the vertical symmetry mid-plane parallel to the arms. The deformation is parabolic along the arm, zero at each arm end, and is parameterized by the maximum downward displacement, Δgap, occurring at the center of the arm.

Unity optical power is injected in either port 1 or port 2, and time-harmonic full-vector 3D Maxwell's equations are solved in the model's volume, which includes all metal, air gaps, and the 100 nm wide empty regions on both sides of the switch, as previously described. For illustration, Figure 4 shows the horizontal plane bisecting the switch through the middle of the gap and plots the instantaneous values of the in-plane magnetic-field component normal to the direction of GP propagation.

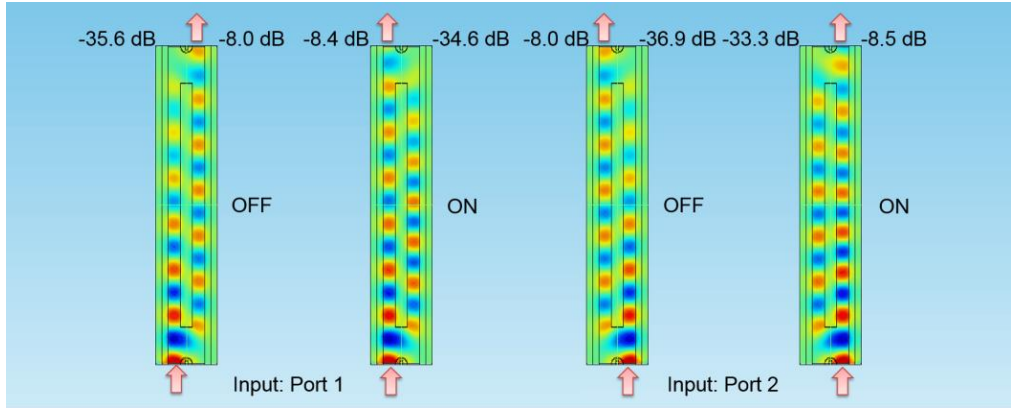

Fig. 4. Instantaneous values of the horizontal magnetic field component normal to the direction of GP propagation, in the mid-plane of the gap in the switch ON ($\Delta gap$ = 4.6 nm) and OFF states under port 1 and port 2 excitation. Transmission to each of the output ports is marked for each case. The simulated worst case insertion loss and extinction ratio is -8.5 dB and 25.3 dB respectively.

Solving the model and calculating the switch S-parameters for each of the two excitation ports and for a series of $\Delta gap$ values (Figure 5) gives the following results: after optimization of the 3 dB coupler, in the OFF state, there is a modest insertion loss of $S41 = S32 = -8$ dB, and low crosstalk of $S41 - S31 = 27.6$ dB and $S32 - S42 = 28.9$ dB. The ON state is reached at $\Delta gap = 4.6$ nm, where the total loss $S31 = -8.4$ dB and $S42 = -8.5$ dB is achieved. The crosstalk in the ON state is low, since $S31-S41 = 26.2$ dB and $S42-S32 = 24.8$ dB. By comparing the ON and OFF states for each input and output port combination, it is evident that the worst case extinction ratio is $S32_{OFF} - S32_{ON} = 25.3$ dB. The switching $\Delta gap = 4.6$ nm is in good agreement with the expected $0.28 \cdot g_0 = 4.76$ nm from the initial parameter estimates based on [1].

We note that in the ON state, ports 1 and 2 are not, in principle, equivalent, because only one arm of the switch is deformed, but this asymmetry is small. Finally, the reflections ($S11$ and $S22$) and input-port cross-coupling $S21 = S12$ are below -17 dB for both switch states.

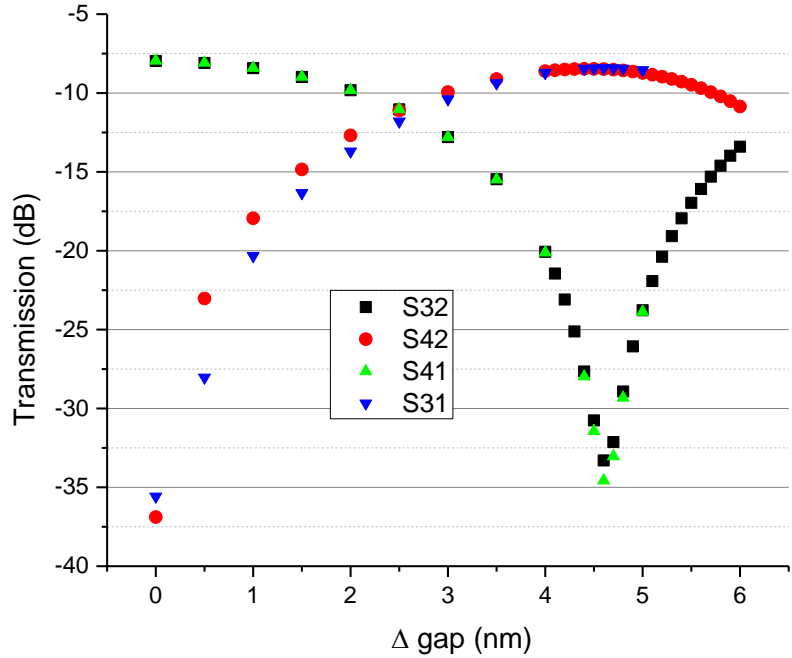

Fig. 5. Transmission to each of the output ports from each or the input ports as a function of the gap change. Switch operation is nearly symmetric w.r.t. input port.

Based on a 1/e (4.3 dB) insertion loss through the 2 μm phase modulator [1], the lower loss limit for a 2.5 μm MIM is about 5.4 dB, due to losses in the gold. It is likely that the additional 2.6 dB to 3.1 dB loss in the modeled switch results from scattering in the transition regions between the ports and the coupler, and the coupler and the modulator arms. Further loss improvement may be achieved there.

**Conclusion**

A small nanomechanical gap-plasmon 2x2 switch design is presented with performance numerically verified. The device takes advantage of the tight confinement and strong opto-mechanical phase modulation possible in MIM structures. While the gap is only 17 nm, a reasonable 8.5 dB insertion loss is made possible by the short device length of only 2.5 μm. The switch footprint is only 0.5 μm wide for a total area of 1.25 μm$^2$.

**Acknowledgement**

The author would like to thank Brian Dennis and Girsh Blumberg for their comments on the manuscript, and is particularly grateful to Brian Dennis for his help in manuscript preparation.